# On the theoretical RF field limits of multilayer coating structures of superconducting resonator cavities.

Alex Gurevich,
Department of Physics, Old Dominion University, Norfolk, VA, 23529, USA

**Abstract**

This Comment addresses theoretical field limits for superconducting-insulating (S-I) thin films multilayers discussed by S. Posen, G. Catelani, M. Liepe, J. Sethna, and M. Transtrum [1]. It is shown that their criticism of the SIS multilayer approach [2] to reduce vortex dissipation in superconducting RF resonator cavities is unsubstantiated.

**Introduction**

Recently S. Posen, G. Catelani, M. Liepe, J. Sethna, and M. Transtrum have discussed theoretical field onsets of penetration of vortices in superconducting-insulator (SIS) multilayers under dc magnetic fields. The authors of Ref. 1 concluded that thin film SIS multilayer structures deposited on the inner surface of superconducting radio-frequency (SRF) resonator cavities suggested in Ref. 2 do not significantly increase the field onset of penetration of vortices and result in extremely high dissipation. The main points of Ref. 1 can be summarized as follows.
1. The lower critical field $H_{c1}$ of SIS multilayers is zero so they do not protect the SRF cavities against magnetic flux penetration by enhancing $H_{c1}$ in thin layers, which according to Ref. 1, was the main point of Ref. 2.
2. SIS multilayers do not significantly enhance the superheating field $H_{sh}$ as compared to Nb even if the S layers are made of materials (like $Nb_3Sn$) with the thermodynamic critical field $H_c$ much higher than $H_c$ of Nb.
3. Thin film SIS multilayers exhibit extremely high rf dissipation due to penetration of vortices. According to Ref. 1, dissipation in SIS multilayers is much stronger than in a thick $Nb_3Sn$ film deposited on the inner surface of a Nb cavity.

In this Comment I show that these statements are incorrect because they result from either inadequate physical assumptions or misinterpretation of Ref. 2.

### 1. $H_{c1}$ in thin film SIS multilayers

The authors of Ref. 1 calculated the thermodynamic potential G(x) of a single vortex as a function its position x across a SIS structure, using the London equation in which the superfluid density $n_s$ is assumed constant everywhere but the vortex core replaced with a rigid cylindrical region of radius $r_0 \sim \xi$ in which the circulating currents discontinuously drop to zero. Here the superconducting coherence length $\xi$ is assumed to be much shorter than the London penetration depth $\lambda$ and the thickness d of S film. In the London model the exact value of the core cutoff radius $r_0$ remains undetermined because the model ignores the current pairbreaking effects resulting in a gradual decrease of $n_s(r)$ in the vortex core in more consistent Ginzburg-Landau (GL) or Eilenberger theories. Except for the small core region, the London model adequately describes circulating currents of vortices in films of type-II superconductors with $\kappa_{GL} = \lambda/\xi \gg 1$ and thickness $d \gg \xi$. However, the principal uncertainty of the London vortex core cutoff is essential for the understanding of the well-known limitations and artifacts of the London model as the distance of the vortex core from

I layer becomes smaller than ~ $\xi$. Disregarding these limitations in Ref. 1 resulted in the wrong calculation of $H_{sh}$, which will be addressed in the next section.

Using the London model, the authors of Ref. 1 reproduced many old results on vortices in films in uniform magnetic fields H applied at both sides of the film or in superconducting screens in which the filed is applied at one side of the film [3-5]. Only the last case is relevant to the SRF cavities, and it is the case for which the SIS structures were analyzed in Ref. 2. For a S-I bilayer on top of a semi-infinite superconductor, G(x) formally vanishes as the vortex reaches I layer. The latter occurs at zero applied field H, so the authors of Ref. 1 concluded that there is a stable localized vortex position at the I layer, and the SIS multilayers have $H_{c1}$ = 0, which allegedly contradicts Ref. 2. However, neither the stable position of a vortex nor the vortex itself in I layer exists: as the vortex gets close to I layer it jumps into it due to strong attraction to the anti-vortex image (similar to the vortex at the surface). As a result, the normal core disappears and the vortex flux quantum $\phi_0$ spreads over the entire I layer, resulting in antiparallel Meissner screening currents flowing at both sides of I layer. In this case $H_{c1}$= 0 defined formally in Ref. 1 does not mean any enhanced vortex dissipation or spontaneous generation of vortices usually associated with a vortex state with $H_{c1}$=0. Instead the state with $H_{c1}$=0 in Ref. 1 is just a trapped flux in the I layer which does not cause any extra rf dissipation because it does not have the main feature, which causes strong dissipation of vortices: the normal cores oscillating under the rf field. The flux trapping can be caused by any weak unscreened stray fields penetrating into I layer from the edges, as shown in Fig. 1a. This effect does not cause dissipation in the SRF cavities and can be eliminated in the cavity geometry in which the edges of the SIS layers are protected from magnetic fields inside the cavity. Alternatively, the SIS multilayer can cover only the equatorial part of the cavity where H is maximum, but the edges of I layer are protected as depicted in Fig. 1b. Here magnetic flux can only penetrate into I layer due to vortices driven by Meissner currents across S layer.

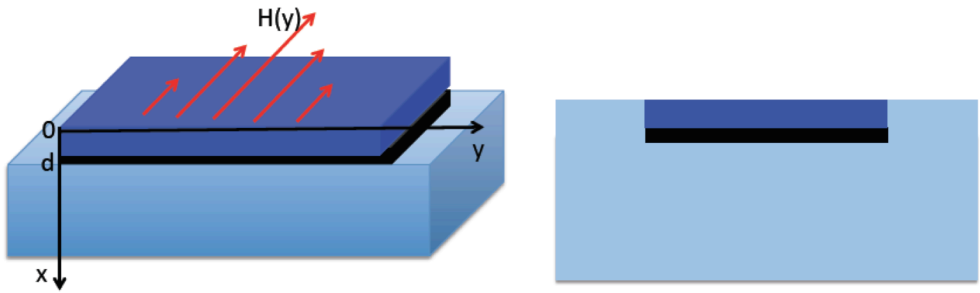

**Fig. 1.** (a) A thin film S-I bilayer on top of a bulk superconductor in a parallel nonuniform magnetic field H(y). The dark blue shows a high-$H_c$ superconducting (S) film of thickness d separated from the bulk Nb substrate (light blue) by insulating (I) layer (black). Unscreened stray fields can leak into I layer from the edges (left panel). (b). Geometry of the S-I bilayer in which I layer is protected from the edge field penetration (right panel).

The goal of the SIS structures of Ref. 2 was to reduce (ideally to eliminate) strong dissipation of vortices in a geometry for which no thermodynamically stable position of a single vortex exists up to H ≈ $H_{sh}$. As was shown in Ref. 2, this could be done by depositing a multilayer of thin (d < $\lambda$) superconducting (S) layers with high $H_c$ on the inner surface of a Nb cavity. Here S layers are decoupled by dielectric layers which suppress the Josephson currents between the screening S layers so that H(x) at the interface with the thick Nb substrate becomes smaller than the bulk $H_{c1}$ ≅ 150 mT of a clean Nb. The formula for the enhanced parallel lower critical field $H_{c1}^{(f)}$ = $(2\phi_0/\pi d^2)[\ln(d/\xi) - 0.07]$ was given in Ref. 2 to illustrate the well-known feature of thin films in a uniform magnetic field equal at both sides of the film [3], but not to justify the SIS multilayer approach. In fact, the enhanced $H_{c1}^{(f)}$ has little relevance to the screening SIS structures, except for the special metastable state in which the trapped field in I layer produces the magnetic field equal to the applied field H. However, in thicker films with $d \gg \lambda$, a thermodynamically stable vortex position appears at $H > H_{c1}^{(f)}(d)$ and strong rf vortex dissipation is restored (see below).

## 2. Superheating field in SIS multilayers

It was stated in Ref. 1 that the SIS multilayers do not really increase the superheating field $H_{sh}$. This conclusion was made from the evaluation of the energy barrier for the vortex entry in the London theory, which principally cannot be used to calculate $H_{sh}$. Indeed, $H_{sh}$, by definition, is the field at which the vortex-free Meissner state becomes absolutely unstable with respect to infinitesimal perturbations of the superconducting order parameter and the magnetic vector potential, so the London theory in which the superfluid density is assumed constant is inadequate. Posen et al. evaluated the field at which the barrier for the vortex entry disappears, assuming that the crude London approximation of the rigid vortex core with undetermined cutoff can be used for the vortex sitting at the S-I interface. This assumption contradicts numerous numerical simulations of the GL equations [6-8] which have shown that the order parameter and circulating currents in the vortex crossing the S-I interface change radically as compared to the London model which fails if the distance between the vortex core and the S-I interface becomes of the order of $\xi$. The London model can only be used for rough qualitative estimates of $H_{sh}$, but the conclusions of Ref. 1 about a few % increase of $H_{sh}$ are not theoretically substantiated.

It is unclear why the London model was chosen for the evaluation of $H_{sh}$ in the first place, given the existence of the well-developed theory for the calculation of $H_{sh}$ using Ginzburg-Landau [9,10], BCS or Eilenberger equations [11-13], to which some of the co-authors of Ref. 1 have contributed as well. For a thin film SIS multilayer, the Meissner screening current is nearly uniform across the S-layers, and the $H_{sh}$ calculated from the stability analysis of the BCS or Eilenberger equations at $T \ll T_c$ without the unjustified London assumptions gives the simple asymptotically exact result for $\kappa_{GL} \gg 1$ [11-13]:

$$H_{sh} \approx 0.84 H_c \qquad (1)$$

Here $H_c$ is the thermodynamic critical field of S layers. Impurities affect $H_{sh}$ weakly [13], while the small thickness of S layers may also slightly increase $H_{sh}$ by suppressing the finite-k instability [9]. The estimate of $H_{sh}$ for $Nb_3Sn$ ($H_c$ = 540 mT) from Eq. (1) gives $H_{sh}$ = 454 mT, or $H_{sh}$ = 406 mT for the numbers from Table 1. The so-obtained $H_{sh}$ is nearly twice the superheating field of Nb and since $H_{sh}$ is also the field at which the entry barrier for vortex penetration vanishes, the use of SIS multilayers offers an opportunity to extend the weakly-dissipative Meissner state to much higher rf fields as compared to the Nb cavities [2].

## 3. Vortex dissipation in SIS multilayers

It was stated in Ref. 1 that SIS multilayers exhibit very high level of dissipation due to penetration of vortices. This point was illustrated by taking Eq. (10) of Ref. 2 for the power P(H) caused by penetration of vortices at $H = H_{sh}$. However Eq. (10) of Ref 2 was derived for global penetration of vortices of the high areal density $\sim H_{sh}/\phi_0$ so this P(H) is of the order of the power released during the quench transition of the entire SRF cavity from the Meissner to the normal state. This power is indeed "unimaginably high for SRF applications" [1] but it also has no relevance to the rf power characteristic of stable operation of SRF cavities. Comparing P(H) released during the global penetration of vortices into SIS multilayers with typical P(H) of SRF cavities Posen et al. suggested that thick $Nb_3Sn$ films deposited onto Nb cavities would be more beneficial for boosting the SRF performance at high fields.

The conclusion of Ref. 1 is based on incorrect comparison of multilayers and thick films under very different conditions. To make this comparison more consistent, I calculate the energies released during penetration of a parallel vortex into a SIS multilayer ($W_{ml}$) and a thick film ($W_{tf}$) under the same conditions. The energy W per unit length of the vortex is the work of the Lorentz force to move the vortex from point x = 0 to x = $x_m$:

$$W = \phi_0 \int Jv dt = \phi_0 \int_0^{x_m} J(x,t) dx = \phi_0 [H(0) - H(x_m)] \quad (2)$$

where v = dx/dt is the vortex velocity, J(x,t) = - dH/dx is the current density defined by the screened magnetic field H(x) = Hexp(-x/λ). Here $x_m$ >> λ for a thick film and $x_m$ = d < λ for a multilayer in which the vortex gets intercepted by the first I layer and turns into non-dissipative trapped flux, as discussed above. From Eq. (2), it follows that:

$$W_{tf} = \phi_0 H, \qquad W_{ml} = \phi_0 H [1 - e^{-d/\lambda}] \simeq (d/\lambda) W_{tf} \quad (3)$$

Therefore, contrary to the above statement of Ref. 1, the same elementary act of penetration of a single vortex produces (d/λ) < 1 less dissipated energy in a SIS multilayer than in a thick film [2] because the I layers impede propagation of vortices. This also follows from Eq. (10) of Ref. 2 in which P(H) is reduced by the factor d/λ as compared to the corresponding P(H) for thick films. More complicated cases of dissipation of vortices under the rf field were addressed in Ref. [14].

### 4. Comparison of thick films with SIS multilayers

To see if thick films could be better than SIS multilayers, it is instructive to compare multilayers and thick films from a broader perspective of minimizing the vortex dissipation. The fact that thin film SIS multilayers provide no equilibrium position of a single vortex in the entire sample up to H ≈ $H_{sh}$, is certainly beneficial for reducing RF dissipation. By contrast, a thick film with d >> λ under magnetic fields higher than the bulk lower critical field, H > $H_{c1}$ = $(\phi_0/4\pi\mu_0\lambda^2)[\ln(\lambda/\xi)+0.5]$ [15] would have an equilibrium vortex lattice spaced by a = $(\phi_0/B)^{1/2}$ < λ from the surface [16] where B = $\mu_0$H. Such equilibrium vortex structures produce much stronger rf dissipation than the exponentially low BCS dissipation in the Meissner state. Thus, the use of thick type-II superconducting coatings with high-$H_c$ but low $H_{c1}$, would assume that many SRF cavities could somehow reliably operate in a highly metastable state at H > $H_{c1}$ being protected by only the Bean-Livingston surface barrier. The assumption of Ref. 1 that "enhancement of $H_{c1}$ is not necessary" disregards a high statistical probability that many common surface defects in SRF cavities can locally reduce this barrier [17,18], triggering penetration of highly dissipative dendritic vortex avalanches which would cause strong Q slope and cavity quench at fields ~ $H_{c1}$ ≈ 50 mT.

Another issue with thick films is that high-$H_c$ type-II superconductors usually have much lower thermal conductivity κ than Nb, so depositing film structures on the inner cavity surface increases thermal resistance of the cavity wall, causing overheating effects. Because layers of the SIS structure have mush smaller thickness $d_i$ = 1-2 nm and $d_s$ = 30-50 nm than 1μm thick $Nb_3Sn$ film mentioned in Ref. 1, SIS multilayers would increase the thermal resistance to a lesser extent than thick films. For a SIS multilayer consisting of

Nb$_3$Sn with $\kappa_s$ = 10$^{-2}$ W/mK [19] and the total thickness d$_s$ = 100 nm, and Al$_2$O$_3$ with $\kappa_i$ = 0.3W/mK [20] and the total thickness d$_i$ = 10nm at 2K, the thermal conductivity is dominated by phonons with wavelengths ~ 100 nm much larger than the thickness of I layers. In this case I layers do not impede heat diffusion across the multilayer which causes the extra thermal resistance, G$_{ml}$ = d$_s$/$\kappa_s$ + d$_i$/$\kappa_i$ = 10$^{-5}$ W/m$^2$, which is about 10% of the thermal resistance G$_{Nb}$ = d$_{Nb}$/$\kappa_{Nb}$ = 10$^{-4}$ W/m$^2$ of the Nb cavity wall for $\kappa_{Nb}$ = 20 W/mK and d$_{Nb}$ = 2 mm. However, a 1 $\mu$m thick Nb$_3$Sn film deposited onto the Nb cavity wall doubles its thermal resistance, which may cause stronger overheating effects (amplified by stronger vortex dissipation) and significantly reduce the thermal breakdown field [21].

## 5. Conclusions

This comment shows that none of the points of Ref. 1 about disadvantages of SIS multilayers were substantiated, while the statement about potential advantages of thick Nb$_3$Sn films was not supported by any theoretical analysis or experimental data. Thick Nb$_3$Sn films are indeed easier to deposit onto the Nb cavities but the above analysis indicates that these films would be prone to stronger dissipation of vortices and overheating effects at B > B$_{c1}$ $\approx$ 50 mT. From the theoretical point of view, SIS multilayers may provide the most effective protection against dissipative penetration of vortices.